\input harvmac
\input epsf

\vskip 1cm

 \Title{ \vbox{\baselineskip12pt\hbox{  Brown Het-1229 }}}
 {\vbox{
\centerline{  Hidden classical symmetry in quantum spaces  }
\centerline{   at roots of unity : From $q$-sphere to fuzzy sphere.  }  }}

\def\cH{ { \cal H} }
\def\Af{ { \cal  A}_f }
\def\Aq{ { \cal  A}_q }

\centerline{$\quad$ { Antal Jevicki, Mihail Mihailescu
 and Sanjaye Ramgoolam }}
\smallskip
\centerline{{\sl  }}
\centerline{{\sl Brown  University}}
\centerline{{\sl Providence, RI 02912 }}
\centerline{{\tt antal,mm,ramgosk@het.brown.edu}}
 \vskip .3in 

 We study relations between different kinds of non-commutative spheres
 which have  appeared in the context of ADS/CFT correspondences 
 recently, emphasizing the connections between spaces 
 that have manifest quantum group symmetry and spaces 
 that have manifest classical symmetry. In particular
 we consider  the quotient $SU_q(2)/U(1)$ at roots of unity, 
 and find its relations with the fuzzy sphere with 
 manifest  classical $SU(2)$ symmetry. 
 Deformation maps between classical and quantum symmetry,
 the $U_q(SU(2))$ module structure of quantum spheres
 and the structure of indecomposable representations
  of $U_q(SU(2))$ at roots of unity conspire in an interesting way to
 allow  the relation between  manifestly $U_q(SU(2)$ symmetric 
 spheres and manifestly $U(SU(2))$ symmetric spheres. The relation
 suggests that a subset of field theory actions on the q-sphere are
 equivalent to actions on the fuzzy sphere. 
 The results  here are compatible with the proposal that 
 quantum spheres at roots  of unity appear as effective 
 geometries which account for finite $N$ effects
 in the ADS/CFT correspondence.

\lref\jr{ A. Jevicki and S. Ramgoolam, ``Non-commutative gravity 
 from the ADS/CFT correspondence,'' { \it JHEP}  { \bf 9904 } 
(1999) 032, hep-th/9902059.  } 
\lref\sus{ J. Mc Greevy, L. Susskind, N. Toumbas,
     ``Invasion of the Giant Gravitons from Anti-de Sitter Space,'' 
       {\it JHEP}  { \bf 0006} 008, 2000, hep-th/0003075 }
\lref\myers{ R.C. Myers, ``Dielectric-Branes,'' { \it JHEP}  {\bf
 9912}  (1999) 022, 
hep-th/9910053.   } 
\lref\hrt{ P.M.Ho, S.Ramgoolam, R.Tatar, 
 ``Quantum Spacetimes and Finite N Effects in 4D Super Yang-Mills
 Theories,'' { \it Nucl. Phys. } 
 {\bf B573 }  (2000) 364-376, hep-th/9907145 }
\lref\berve{ M. Berkooz and H. Verlinde,  ``Matrix Theory, AdS/CFT and
 the Higgs-Coulomb equivalence,'' hep-th/9907100, { \it JHEP } 
{\bf 9911} (1999) 037 }
\lref\mmp{S. Meljanac, M. Milekovic, S. Pallua, hepth/9404039 }  
\lref\pmt{ Pei Ming Ho, ``Geometrical aspects of quantum spaces,''
 PhD thesis, UMI-97-03152-mc }  
\lref\madore{J. Madore, ``The fuzzy sphere,'' { \it
 Class. Quant. Grav. }  
{\bf 9}  (1992 ) 69-87.  } 
\lref\bigq{ ``Quantum 2-spheres and big-q-Jacobi Polynomials,''  
             M. Noumi and K. Mimachi, CMP 128, 521-531 (1990).   } 
\lref\keller{G. Keller, `Fusion rules of $U_q (sl(2,C)), q^m =1$''
            { \it Lett. Math. Phys. }  {\bf 21}  :273-286, 1991  }
\lref\stein{ H. Steinacker, J. Madore, H. Grosse, 
   ``Field Theory on q-deformed fuzzy sphere I.''  hepth/0005273. } 
\lref\bwsft{ B. Zwiebach and E. Witten , ``Two dimensional String 
    Field theory,'' }
\lref\pmli{ Pei Ming Ho  and Miao Li,  ``Fuzzy spheres from the ADS/CFT 
correspondence  and  holography from non-commutativity,''
 hep-th/0004072 } 
\lref\gerst{P. Bonneau, M. Flato, M. Gerstenhaber, G. Pinczon, 
{ \it Commun. Math. Phys. }  {\bf 161 }, (1994) 125 }  
\lref\cgz{ T. L. Curtright, G. I. Ghandour and C. K. Zachos, 
 ``Quantum algebra deforming maps, Clebsch-Gordan coefficients,
 co-products, R and U matrices, ''{ \it  J. Math. Phys. } 
 {\bf 32 } (1982) 3 } 
\lref\bv{ M. Berkooz and H. Verlinde, 
``Matrix Theory, AdS/CFT And Higgs-Coulomb Equivalence, { \it JHEP }  
{\bf 9911} :037,1999,   hep-th/9907100  } 
\lref\malstrom{ J. Maldacena, A. Strominger, ``ADS3 black holes 
                   and a Stringy Exclusion Principle,''
{ \it JHEP}  9812 (1998) 005 hepth/9804085.   } 
\lref\jdm{ S. Das, A. Jevicki and S.  Mathur 
``Giant gravitons, BPS bounds and noncommutativity,'' hep-th/0008088.  } 
\lref\mytf{M. Grisaru, R. Myers, O. Tafjord, `` SUSY and Goliath,'' 
 hep-th/0008015 } 
\lref\hhi{ A. Hashimoto, S. Hirano, N. Itzhaki, 
 ``Large branes in AdS and their field theory dual,'' hepth/000816 }  
\lref\jmri{ A. Jevicki, M. Mihailescu, S. Ramgoolam, 
``Gravity from CFT on $S^N (X)$ : Symmetries and interactions''
  { \it Nucl.Phys.} { \bf B577} :47-72,2000,  hep-th/9907144. } 
\lref\jmrii{ A. Jevicki, M. Mihailescu, S. Ramgoolam, 
`` Noncommutative spheres and the AdS/CFT correspondence,''
hep-th/0006239. } 
\lref\matlun{ S. Mathur, O. Lunin, 
``Correlation functions for $M^N/S_N$ orbifolds,''
 hep-th/0006196 }
\lref\sugi{ T. Sugitani,  `Harmonic analysis on quantum spheres
 associated with representations of $U_q (so_N)$ and q-Jacobi 
 Polynomials,'' { \it Compositio. Math. }  {\bf 99}; 249-281, 1995 }  
\lref\pmstu{ P.M. Ho and Yu Ting Yeh, `Non-commutative D-brane in
 non-constant NS-NS B-field background,'' hepth/0005159 } 
\lref\duff{ M. Duff, H. Lu and C. N. Pope, ``$AdS3 \times S^3$ (un)twisted
 and squashed and an $O(2,2,Z)$ multiplet of dyonic strings,''
 { \it Nucl.Phys.} { \bf B544} (1999) 145, hep-th/9807173  } 
\lref\drinf{ V. G. Drinfeld, { \it Leningrad. Math. J. } 
 {\bf 1}, 1419 (1990)  } 
\lref\cqri{ R. Coquereaux, A. O. Garcia, R. Trinchero, 
``Finite dimensional quantum group covariant differential calculus on
 a complex matrix algebra,'' math.QA/9804021 } 
\lref\cqrii{ R. Coquereaux, G. E. Schieber, ``Action of finite quantum
 group on the algebra of complex $N \times N $ matrices,'' math-ph/980716 }
\lref\wzwqg{ G. Moore and N. Seiberg, 
{\it  Commun. Math. Phys. } 123, 1989, 177
 \hfill\break
V. Pasquier and H. Saleur,
 { \it Nucl.Phys. } { \bf B330 }, (1990) 523. \hfill\break 
L.Alvarez-Gaume, C.Gomez and G.Sierra,{\it  Nucl.Phys.} {\bf B319}
(1989)155; \hfill\break
 A. Alekseev and S. Shatashvili, {\it  Commun. Math. Phys.} {\bf 133}
(1990) 353. } 
\lref\csw{ J. Castellino, S. M. Lee, W. Taylor, ``Longitudinal
5-branes as four-spheres in Matrix Theory,'' { \it Nucl. Phys. } { \bf B526}:
334-350, 1998 } 
\lref\steinck{ H. Steinacker, ``Finite dimensional unitary
 representations  of quantum Anti-de Sitter groups at roots of
 unity,''q-alg/9611009, { \it Commun.Math.Phys. }  {\bf 192}  (1998) 687 } 
 \lref\suz{ T.  Matsuzaki and T. Suzuki, 
`` Unitary highest weight representation of $U_q(SU(1,1))$ 
                 when $q$ is a root of unity,''
                { \it J. Phys. A : Math. Gen.}  { \bf } 26 (1993) 4355-4369 } 
\lref\roch{E. Buffenoir and 
P. Roche ``Harmonic analysis on the quantum Lorentz group,''
{\it  Commun. Math. Phys.}  {\bf 207 } 499-555, 1999. } 
\lref\tesh{B. Ponsot and 
 J. Teschner ``Clebsch-Gordan and Racah-Wigner Coefficients for a
 continuous series of representations of $U(Q)(SL(2,R))$, ''
math.qa/0007097 }
\lref\liho{ P.M.Ho and M. Li
``Large N Expansion from fuzzy ADS2,'' hep-th/0005268 }
 \lref\recfuz{  A. Alekseev, A. Recknagel, V. Schomerus, 
``Noncommutative World Volume Geometries: Branes on $SU(2)$ 
and fuzzy spheres.'' { \it JHEP}  {\bf 9909}:023,1999,
 hep-th/9908040. \hfill\break
N. R. Constable, R. C. Myers, O. Tafjord, ``The non-commutative Bion
 core,'' Phys.Rev. D61 (2000) 106009, hep-th/9911136 \hfill\break
C.V. Johnson, ``Enhancons, fuzzy spheres 
 and multi-monopoles,'' hep-th/0004068. \hfill\break  
D. Berenstein, V. Jejjala, R. Leigh, ``Marginal and
 relevant deformations of $N=4$ field theories and Non-commutative
 moduli spaces, '' hep-th/0005087 \hfill\break
K. Dasgupta, S. Hyun, K. Oh, R. Tatar,`` 
Conifolds with Discrete Torsion and Noncommutativity,'' hep-th/0008091
\hfill\break
J. Polchinski, M. Strassler, ``The string dual of a
 confining 4-dimensional gauge theory,'' hepth/0003136   \hfill\break 
S. Trivedi, S. Vaidya, ``Fuzzy cosets and their gravity
 duals,'' hep-th/0007011   \hfill\break
C. Bachas, J. Hoppe, B. Pioline, 
``Nahm equations, $N=1*$ domain walls, and D strings in 
$ADS(5) X S(5)$,''  hep-th/0007067   } 
\lref\wat{P. Watts, ``Derivatives and the role of the Drinfeld twist
 in Non-commutative string theory,'' hepth/0003234 }
\lref\gz{ A. Giaquinto and J. Zhang, ``Bialgebra actions, twists, and 
 universal defmroation formulas, '' hep-th/9411140 }  
\lref\cc{C. Chryssomalakos,
 ``Drinfeld twist for quantum $su(2)$ in the adjoint representation,''
 { \it Modern Phys. Letts. A,}  Vol. 13, no. 27 (1998) 2213 }   
\lref\hsi{ H. Steinacker ``Quantum Anti-de-Sitter space and sphere at roots
 of unity,'' hepth/9910037 } 
\lref\chpr{ V. Chari and A. Pressley, ``A guide to quantum groups,''
 CUP 1994 }  
\lref\yoney{ T. Yoneya, ``String Theory and the Space-Time Uncertainty
principle'' hep-th/0004074 } 
\lref\podles{ P. Podles, ``Quantum spheres,'' { \it
 Lett. Math. Phys. }  { \bf 14 } (1987) 193-202. }  
\lref\sw{ N. Seiberg and E. Witten, ``
String Theory and Noncommutative Geometry,'' { \it JHEP }  9909 (1999) 032,  
hep-th/9908142 }
\lref\jeyo{ A. Jevicki and T. Yoneya, ``Spacetime Uncertainty
 principle and Conformal symmetry in D-particle dynamics,''
 { \it Nucl. Phys. } { \bf  B535 }  ( 1998 ) 335, hepth/9805069 } 
\lref\polyk{ A. Polykronakos,  ``A classical realization of quantum
 algebras,'' {\it Mod.Phys.Lett.}  { \bf A5}: 1990, 2325. }  
\lref\oeck{ R. Oeckl, 
``Untwisting noncommutative $R^d$ and the equivalence
of quantum field theories,'' 
{ \it Nucl.Phys.B} { \bf  581 }  (2000) 559-574, hep-th/0003018\hfill\break
``Braided Quantum Field Theory'', hep-th/9906225 \hfill\break
``The Quantum Geometry of Spin and
 Statistics,'' hep-th/0008072 } 
\lref\dob{ V.K. Dobrev and P.J. Moylan, ``Finite-dimensional singletons of
the quantum anti de Sitter algebra,'' { \it Phys. Lett.} {\bf 315B} (1993)
292-298 \hfill\break 
V.K. Dobrev and R. Floreanini, The massless representations
of the conformal quantum algebra, { \it J. Phys. A: Math. Gen.}  {\bf 27}
(1994)  4831-4840. }

%\draftmode
\Date{ July 2000 }

\newsec{ Introduction } 

 Non-commutative  spacetimes which are deformations 
 $ADS \times S$ backgrounds 
 have been studied as space-time explanation of the 
 stringy exclusion principle \malstrom, beginning in \jr\ 
 and  further in \hrt\jmri\hsi\bv\pmli, 
 Another mechanism  emphasizing non-commutativity
 uses the  fuzzy sphere world-volumes 
 of fat gravitons 
 moving as dipoles on a transverse non-commutative space was given 
 in  \sus, and further explored recently in \hhi\mytf\jdm. 
 The fuzzy sphere worldvolumes appear by the polarization mechanism 
 of Myers \myers. 
 Some qualitative properties of finite $N$ $ ADS \times S$  correlators
 \jmri\matlun\ are reproduced by overlaps of spherical 
 harmonics on the fuzzy sphere \jmrii.

 For odd-dimensional spheres quantum groups
 give the natural non-commutative candidates. 
 For even spheres the candidates discussed 
 so far keep the classical symmetries manifest and are generalizations 
 of the fuzzy sphere \madore. One motivation of this 
 paper is to begin a study of the relation between 
 the non-commutative spaces based on quantum groups 
 and those based on fuzzy sphere and its generalizations. 
  The  goal in this direction 
 is to study the relations, from the point of view 
 of quantum space-time, which are expected to exist in string theory 
 \duff. 
 Another motivation is to better 
 understand the relations between the candidate non-commutative
 spheres appearing as  part of a non-commutative  space-time 
 and the non-commutative structures
 appearing from choosing a splitting of  the spheres into 
 non-commutative world-volume directions of a fat graviton
  and non-commutative
 transverse directions as in \sus.  
 Another purely mathematical motivation is to ask if there is a
 generalization  to the world of non-commutative spheres 
 of  relations of the kind  $ S^2 = S^3/ U(1) = SU(2)/U(1)$. 
 
 The important  feature that has to emerge from any 
 convincing step in this direction is to clarify in what sense 
 having a quantum group symmetry is compatible with the classical
 symmetries. Uncovering quantum group symmetry in physical systems 
 with manifest classical symmetry has been undertaken 
 in the context of WZW-quantum group correspondences
 \wzwqg. Some mathematical work in the direction of uncovering 
 classical symmetry in quantum groups has also been done, 
 see for example \cgz\polyk\gerst\drinf, and at appropriate points in this 
 paper we will use some of these results.

 In this paper we study 
 the connection between $SU_q(2)/U(1)$ for $q = e^{ i \pi \over M } $
 and fuzzy sphere generated by $SU(2)$ generators  satisfying 
 $\sum_i S_i^2 = J(J+1)$.
 The fuzzy sphere algebra decomposes under representations 
 of $SU(2)$ acting iby commutators as irreducible 
 representations of spin $s$ ranging from $0$ to $2J$. 
 $U_q SU(2)$ has reps. which are cutoff at
 $2s \le M-2$ for $q = e^{i \pi \over M}$. This suggests 
 a relation for $ M = 2 ( 2J+1 )   $ between the $q$-sphere 
 and the fuzzy sphere. 

 Section 2 reviews some properties of the fuzzy sphere $\Af$. 
 Section 3 reviews properties of the $q$-sphere $\Aq$ 
 and its $U_q(SU(2)) $ symmetry.  
 Section 4 obtains the decomposition under  $U_q(SU(2))$ of $\Aq$. 
 The decomposition consists entirely of indecomposable representations 
 We observe that this spectrum of indecomposable representations 
 contains a sub-module which is the direct sum of standard 
 representations of the kind appearing in the decomposition of the fuzzy
 sphere. This allows us in section 5, to develop, 
 using deformation maps \cgz, the precise relations between $ \Aq$ and $\Af$.

\newsec{ Fuzzy sphere } 

 The fuzzy sphere is defined as the algebra 
 generated by the three generators $S_3, S_+, S_-$ 
 obeying the relations of the $SU(2)$ Lie algebra : 
\eqn\surels{\eqalign{& [ S_{+}, S_{-} ] = 2S_3 \cr 
                     & [S_3, S_{+} ] = S_+ \cr 
                     & [ S_3,S_-] = -S_{-}. \cr }}
together with a constraint on the Casimir : 
\eqn\cas{ S_3^2 + { 1\over 2} (S_{+}S_{-}+   S_{-}S_{+} ) = J(J+1) } 
This algebra is infinite dimensional. For example 
 $S_-^l$ for any $l$  are independent elements. 
 It admits however a finite dimensional quotient
 which is isomorphic to the algebra of $ N \times N$ 
 matrices where $N = 2J + 1 $. We will call this finite
 dimensional truncation $\Af ( N) $.  
 
It admits a right action of the universal enveloping algebra 
 of $SU(2)$, by taking commutators from the right.
We could also work with a left action instead  but choose 
   to work with the right  action for convenience. 
 Under this action of $U(SU(2))$, the $\Af ( N) $
 decomposes as a direct sum of representations
 of integer spin $s$ with unit multiplicity with 
 $s$ ranging over integers $s$ from $1$ to $2J= N-1 $.
\eqn\dec{ \Af = \oplus_{ s=0}^{2J} V_s }

\newsec{  The $q$-sphere } 

 We start with the $q$-deformed algebra of functions  $SU(2)$
 which we call $Fun_q ( SU(2) ) $, 
 generated by $\alpha, \beta, \gamma, \delta $
 which obey 
\eqn\alg{\eqalign{  
 & \alpha \beta = q \beta \alpha ~~ \alpha \gamma =  q \gamma \alpha \cr 
 & \beta \gamma =  \gamma \beta ~~  \beta \delta =  q \delta \beta \cr 
 & \alpha \delta - \delta \alpha = (q - q^{-1} ) \beta \gamma \cr 
 & \alpha \delta - q \beta \gamma = 1  \cr }}
The choice of $SU_q(2)$ real form 
is the choice of the involution 
\eqn\real{\eqalign{ &  \alpha^*  = \delta \cr
          &  \beta^*   = \gamma \cr  }}
In the last line we have set to $1$ the central element.

 The algebra \alg\ has a left and a right 
 action of $Fun_q ( SU(2) )$. Under the left 
 $U(1)$, the generators $ \alpha, \gamma $ have charge 
 $1$, and the generators $ \beta, \delta $ have 
 charge $-1$. The $U(1)$ invariant sub-algebra is generated 
 by  $\alpha \beta$, $\alpha \delta$ and $\gamma \beta$. 
 For a more complete discussion of 
 the quantum geometry of these q-spheres
 see for example \pmt. 

 Defining the combinations 
\eqn\ealph{\eqalign{&  e_0 = 1 + ( q + q^{-1} ) \beta \gamma, \cr  
&  e_+ = q^{-1} ( q + q^{-1})^{1/2}\alpha \beta, \cr 
&  e_- = - (q + q^{-1})^{1/2} \gamma \delta \cr }}
one finds that we have an algebra belonging to 
 the family of Podles quantum 2-spheres \podles\    
\eqn\deftq{\eqalign{ 
& e_+e_- - e_-e_+ + \lambda e_0^2 = \mu e_0 \cr 
& q e_0 e_+ - q^{-1} e_+e_0 = \mu e_+ \cr 
& q e_-e_0 - q^{-1} e_0 e_- = \mu e_- \cr 
& e_0^2 + qe_-e_+ + q^{-1} e_+e_- = 1,  \cr }} 
where $\lambda = (q - q^{-1})  $, 
 with $\mu = ( q - q^{-1}) $. 

 This algebra has infinite dimensional representations 
 for generic $q$. For roots of unity $q = e^{ i \pi \over M}$ 
 it displays some special properties. It is easy to prove 
 for example that  $e_-^M$ and $e_+^M$ are central 
 elements. We expect that there will be finite dimensional 
 representations of the $q$-sphere algebra for any $\mu$ 
 when $q$ is a root of unity, which can be constructed 
 by the method of highest weights much the same way 
 we construct representations of $U(SU(2))$ or of $U_q SU(2)$. 
 In the following we will focus on the case of the quotient 
 2-sphere. Finite dimensional representations will lead to 
 finite dimensional quotients of the $q$-sphere algebra, much the way
 they do for the fuzzy sphere as discussed in section 2. 
 We will denote these finite dimensional $q$-sphere algebras as
 $ \Aq$.

\subsec{ Finite dimensional truncations of the quotient sphere}

 There are known representations of
 the $Fun_q ( SU(2) ) $ algebra which we will use 
 to obtain representations of its $U(1)$ quotient. By specializing to roots 
 of unity we can obtain finite dimensional truncations 
 of these algebras.  The $Fun_q SU(2)$ algebra has a family of reps. 
 labelled by $t$ \chpr : 
\eqn\repqs{\eqalign{ & \alpha  |k> = ( 1 - q^{2k})^{1/2} |k-1 > \cr    
           & \beta |k > = -q^{k+1} t^{-1} |k> \cr 
           & \gamma |k> =  q^k t |k> \cr 
           & \delta |k> = ( 1 - q^{2k +2} )^{1/2} |k+1 > \cr }}
 Using the expressions  \ealph\ we get a representation 
 of the $S^2_q$ algebra.
\eqn\repqsii{\eqalign{ & e_0 |k > = ( 1 - q^{2k}( 1 + q^2 ) ) |k > \cr 
              & e_+ |k > = -q^k t^{-1} \sqrt{ q + q^{-1} } ( 1 -
 q^{2k} )^{1/2} |k-1 > \cr 
 &  e_- |k> = - t \sqrt{ q + q^{-1} }  ( 1 - q^{2k +2})^{1/2} q^{k+1}
 |k+1 > \cr }}
The parameter $t$ will not affect the form of the 
 finite dimensional quotient algebra $\Aq$, as we will see in explicit
 examples in later sub-sections. 

Specializing to roots of unity, we find finite dimensional 
reps. with $k$ extending from $0$ to $M-1$, since 
$e_+$ annihilates $|M>$ and $e_-$ annihilates $|M-1>$. 
In these finite $M^2$ dimensional reps. $e_0$ 
 van be expressed as a sum of 
\eqn\expeo{ e_0 = \sum_{l=0}^{M-1}  C_l e_-^le_+^l } 
The coefficients can be determined recursively
 by acting successively on $|0>$ ( which determines $C_0$ directly ), 
 and then $|1>$ ( which determines $C_1$ in terms of $C_0$ ), and so
forth. 
We will write down the explicit expressions 
for the cases $M=3,4$ below. 

The first technical result of this paper is to 
give the decomposition of this $M^2$ dimensional 
$U_q$ module algebra in terms of representations 
 of $U_q$. We have explicit proofs for the cases $M=3,4$, 
and we have several tests of the proposed decomposition
in the general case.

\subsec{ Action of $U_q Sl(2)$ } 
  
The right action of the 
  $U_q Sl(2)$ on the $q$-sphere is given below. 
  It can be obtained from  \bigq\ after some changes of variables. 
\eqn\rtact{\eqalign{& 
(e_-)K = q (e_-) ~~   (e_0)K = e_0 ~~ (e_+)K = q^{-1} e_+ \cr 
& (e_-)X_+ = -e_0  ~~ (e_0)X_+ =  ~e_+ ~~ (e_+)X_+ = 0 \cr 
& (e_-) X_- = 0 ~~ (e_0)X_- = -(q +q^{-1}) e_- ~~ 
(e_+)X_- = (q+q^{-1})  e_0 \cr} }  

 We can check that these are indeed consistent 
 with the standard relations of $U_q$, which in our conventions, 
 are : 
 \eqn\relsc{\eqalign{& KX_+K^{-1} = q X_+ ~~~ KX_-K^{-1} = q^{-1}X_- \cr 
 & X_+X_- - X_-X_+ = { (K^2 - K^{-2} ) \over (q - q^{-1}) }  \cr }}
 To obtain the standard form of classical $U(SU(2))$ algebra 
 in the $q \rightarrow 1 $ limit we write $K= q^H$ and get : 
 \eqn\lim{\eqalign{ & [ H, X_+ ] = X_+ \cr 
                    & [ H, X_- ] = - X_- \cr 
                    & [ X_+, X_- ] = 2H \cr }} 
 The $U_q$ algebra admits finite dimensional quotients at roots of
 unity, and acts as finite dimensional symmetry algebras on the 
 $q$-sphere, and as we will show in section 4 on the fuzzy sphere.

 We can also check that the relations 
 of the $q$-sphere are invariant under the action 
 of the $U_q$ symmetry. In checking this 
 we have to use the following action of $U_q$ on products : 
\eqn\prodac{ ( e_ie_j ) X_+ = (e_i)X_+(e_j)K + (e_i)K^{-1}(e_j)X_+ }
  This form of the action on products uses the co-product of the 
  quantum group and the $q$-sphere is a module-algebra for the 
  quantum group. For a general discussion of module 
 algebras acted on by  Hopf algebras we refer the reader to  \chpr. 

 For real $q$ there is a conjugation operation 
 on the $q$-sphere where 
\eqn\conj{\eqalign{& (e_+)^{\dagger} = - ( q + q^{-1} ) e_- \cr 
                   & (e_-)^{\dagger} = - { 1 \over { ( q+q^{-1}) } } e_+ \cr }}
 Using this we can reconstruct the action of $X_-$ from 
 that of $X_+$.  

 Using \rtact\ and \deftq\ we can write down the 
 action of $X_+$ on $e_-^L$ as
\eqn\actxei{\eqalign{ 
( (e_-)^L ) X_+ = - { q^{-L +3 } \over { 1-q^4 } }
 (1-q^{2L}) ( 1 -q^{2(L-1)} )  e_-^{L-1} -
 q^{-L+1}  { (1- q^{4L} ) 
 \over ( 1 - q^4 ) } e_-^{L-1}e_0 }}
Using the conjugations symmetry at real $q$ or 
 directly \rtact\ and \deftq, we can obtain : 
\eqn\actxe{  ( (e_+)^L ) X_- = 
 { q^{-L+2} \over ( 1 -q^2 ) } ( 1 - q^{2L} ) ( 1 - q^{2(L-1)} )
 e_+^{L-1} + q^{-1} { ( 1 - q^{4L} ) \over ( 1 - q^4 ) } ( 1 + q^4) 
  e_0 e_+^{L-1}  } 
 For $q= e^{i \pi\over M}$, the above equations imply  
\eqn\annm{ (e_+^{M}) X_- = 0, ~~~~ ( e_-^M ) X_+ = 0. } 
 This means that the constraints $e_+^{M}=0, e_-^M=0$ 
 are  consistent with the action of the $U_q$ symmetry. 
 The finite dimensional quotient should be a 
 $U_q$ module algebra.  
 As $q \rightarrow 1 $, only the term $e_+^{L-1}e_0 $ survives. 
 When $L = M/2$, the coefficient of $e_+^{L-1}e_0$ vanishes. 
 In this sense this power of $e_+$ shows very non-classical 
 behaviour.

\newsec{ Reduction to Indecomposables } 

 The $q$-sphere module algebra can be decomposed into reps. 
 of $U_q$. We would like to know what kind of reps. appear. 
  We first perform a reduction $ e_+^{M}=0, e_-^{M}=0 $ 
  together with the polynomial expression of
  $e_0$ in terms of $\sum_{l=0}^{M-1} C_l e_-^{l}e_+^{l}$.
 This reduced algebra is spanned by $e_-^{l_1}e_+^{l_2}$, 
 with $0 \le l_1,l_2 \le M-1$, and is therefore $M^2$ 
 dimensional. It contains   $M$ highest weights 
  $ 1, e_-, e_-^2, \cdots e_-^{M-1}$ annihilated by $X_-$
 (this may seem unusual but it is because we are using right 
 action of $U_q$ rather than left action )  
 and $M$ lowest weights $ 1, e_+, e_+^2, \cdots e_+^{M_1}$ annihilated 
 by $X_+$. 
 The following is  a proposed reduction of the 
 module algebra in terms of indecomposables which is 
 consistent with the dimension $M^2$ 
 and with the above set of highest and lowest weights. 
 For even $M$ we propose, 
\eqn\mtwo{  A_{M^2} = \oplus_{k=1}^{ { M \over 2} } I_{0}^{2k}. } 
 We are here using the notation of \keller\ 
 for the indecomposables. 
 Each of these indecomposables has dimension $2M$, 
 so the above is consistent with the 
 dimension count $M^2 = {M\over 2 } (2M)$. 
 For odd $M$ we propose 
\eqn\modtw{ A_{M^2} = \oplus_{k=1}^{ {(M-1)\over 2} }  I_{0}^{2k+1} 
                         \oplus I_0^{1} } 
 Since $I_{0}^{1}$ has dimension $M$ the above is consistent 
 with the dimension count $M^2 = {(M-1) \over 2} (2M) + M$. 
 A similar discussion of quantum group structure of 
 Matrix algebras has appeared in \cqri\cqrii, where the 
 the geometric objects considered were quantum planes rather
 than quantum spheres. 
 
 Counting the number of highest and lowest weights
 gives another check of \mtwo\ and \modtw. 
 Each rep. of the form $  I_{0}^{2k} $ has two highest weight states
 and two lowest weight states. Each rep. of the form 
 $I_{0}^{(1)}$ has one highest weight and one lowest weight. 
 The decomposition in \mtwo\ has $2 ( M/2) = M $ highest 
 weights. The decomposition in \modtw\ has $2 { (M-1) \over 2 } + 1 =
 M$ highest weights. 

 Each representation of the type $I_0^{2k}$ contains as a 
 submodule an ordinary representation of dimension 
 $M-p+1$. 
 The dimensions of these reps add up to $(M/2)^2$ for even $M$, 
 and to $({ M-1\over 2 })^2$for odd $M$. So we can map them 
 to the fuzzy sphere corresponding to the respective matrix 
 algebras while keeping the same structure of 
 $SU(2)$ representations as the ones of classical $SU(2)$. 
 
  The above proposal  for the decomposition 
 of the $A_q$ algebra in terms of indecomposables
 has some implications which can be checked. In the case
 of even $M$ the set of highest weights 
 $e_-^{M/2},  e_-^{M/2+1},   e_-^{M/2+2}, \cdots e_-^{M-1}$
 pair up, respectively, with $e_-^{{M/2}-1}, e_-^{M/2-2}
 \cdots 1$.
 By applying an appropriate number of 
 powers of $X_+$ we get from the upper highest weights
  to the lower ones according to the structure of $I_0^p$ 
 described in \keller. We need to check that  
 \eqn\toch{ (e_-^{ { M - 2 + p }\over 2} )X_+^{p-1} \sim e_-^{k_1-k_2} } 
 for $ p = 2,4, \cdots M$.  
 The equation \actxei\ proves the desired result 
 for $p=2 $, since it shows that in $ ( e_-^{M/2} ) X_+$ 
 only the first term survives. Similarly from \actxe\  $ ( e_+^{M/2} ) X_- $
 is proportional to $e_-^{M/2 -1 } $.
 It will be an interesting exercise to give the explicit 
 proof for other values of $p$. Rather than pursuing this direct route
 for general $M$ and $p$ we give a counting arguments 
 which works in the general case, and we give explicit formulae
 for $q^3=-1$ and $q^4=-1$ in the following sections.

\subsec{ Counting for even $M$. } 

 Let us check that the number of states 
 in the algebra is indeed consistent with the above decomposition. 
 At $H = L$, for positive $L$, the polynomials are spanned
 by 
\eqn\sp{ e_-^{k},  e_-^{k+1}e_+, e_-^{k+2}e_+^2, \cdots
 e_-^{M-1}e_+^{M-1-k} } 
i.e a total of $M-k$ states. 

 For $L <  M/2$, the proposed decomposition
 has representations which contribute states with multiplicity 
 $1$ and representations which contribute states with multiplicity 
 $2$. The representations with $p=2, 4, \cdots ( M - 2k  ) $
 contribute two states each, giving a total 
 of $ M -2L$.  The representations with $p=  M, M-2, \cdots ( M-2L + 2 ) $
 contribute one state each giving a total of 
 $L$ states. Adding these up we get $M-L$ states 
 in agreement with the explicit counting of polynomials. 

 For $H > L/2$ we have representations $I_0^{2l}$ contributing 
 one state each for $ l = L - M/2 +1, L- M/2 + 2, \cdots M-L $, 
 giving us exactly $M-H$ states, in agreement with 
 the counting of polynomials in $e_+,e_-$.  
 
 \subsec{ Counting for odd $M$ } 
 
 In this case, we propose a  decomposition  
  into ${  ( M-1 ) \over 2 } $ repsresentations of dimension 
 $2M$ and one representation of dimension $M$. This spectrum is 
 the set of representations $I_0^{p}$ with $p$ ranging over  
the set  $ p = 1, 3, \cdots M $.  
  The highest weight $e_-^{ {M-1 \over 2 }  } $ belongs to $p=1$. 
  The remaining highest weights pair up 
  as $( e_-^{ {M + p -2 \over 2 }  }, e_-^{ {M - p \over 2 }  }  ) $ 
  in the reps. $ I_0^{p} $, for $p$ ranging 
  over $3,5 \cdots M$. 

 Consider eigenvalues of $H$ which are equal 
 to $L \le { (M-1)\over 2 } -1 $. 
 There are two states with such eigenvalue 
 from all reps $I$ with $(M-p) \ge 2L$, i.e for 
 $p = 3,5, \cdots M-2L$. From these we get a total   
 of $M-1-2L$ states. $I$-reps with $(M-p) < 2L$ give one state each. 
 These values of $p$ are $M-2L +2, M-2L+4, \cdots M$. The 
 representation with $p=1$ also contributes one state, giving a total of
 $H+1$ states coming from representations which contribute one each. 
 Adding up the states from reps which contribute $2$ each we get 
\eqn\add{ (M-2L-1)  + (L+1) = M-L } 
This agrees with the count of independent 
polynomials with $H=L$.

 For eigenvalues $H=L$ which obey 
 $L \ge{ (M-1 ) \over 2 } $ we have $I$-reps 
 contributing one state if $M+p-2 \ge 2L$. 
 This allows $M-L$ different values of $p$, 
 again agreeing with the count of independent 
 expressions of the form $e_-^{l_1}e_+^{l_2}$.

\subsec{ An explicit example : $q= e^{i \pi \over 3 }$ } 

 The expression for $e_0$ in this case is : 
\eqn\ezro{ e_0 = -q^2 -q^2 e_-e_+ + q e_-^2 e_+^2 } 

The highest eigenvalue of $H$ in a rep. 
 $I_0^{p} $ is $j$ which is given by 
$ 2j = M +p -2 $ . 

The structure of the rep. $I_0^1$ is given by
 
\fig\qthr{ 
{Figure 1 }  } 
{\epsfxsize0.35in\epsfbox{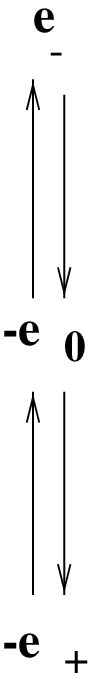} }

The structure of the rep. $I_0^3$ is given below. 

\fig\qthrsec{ 
{Fig.2 }  } 
{\epsfxsize0.85in\epsfbox{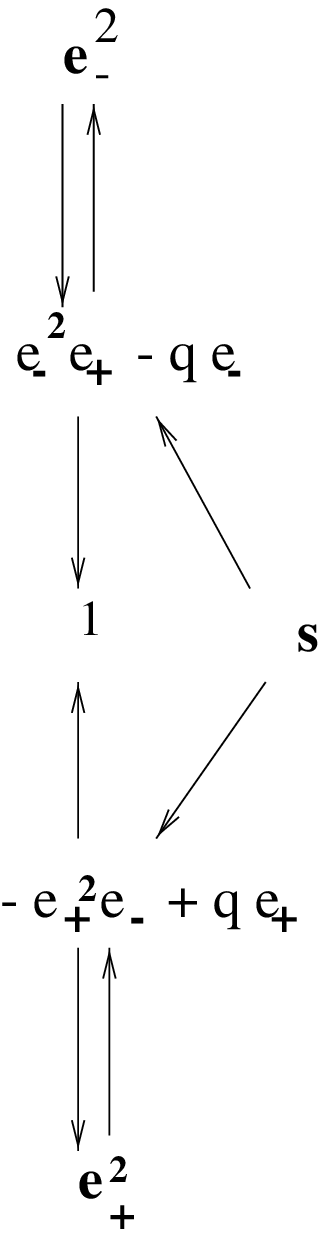} }

The element denoted as $s$ can be determined 
 up to an arbitrary constant to be : 
\eqn\formfs{ s = A_1 + e_-e_+ + q ( 1+ q ) e_-^2e_+^2 } 

\subsec{ An explicit example : $q = e^{ i \pi \over 4 }$ } 

 \eqn\ezroi{ e_0 = -q^2 - q^3 e_-e_+ + {(q^2-1)\over 2 }e_-^2e_+^2 - 
     { q(1 + q^2)\over 2 } e_-^3e_+^3    } 

The rep. $I_0^2$ is shown below. 
\bigskip
\fig\qfr{ 
{Fig.3 }  } 
{\epsfxsize1.0in\epsfbox{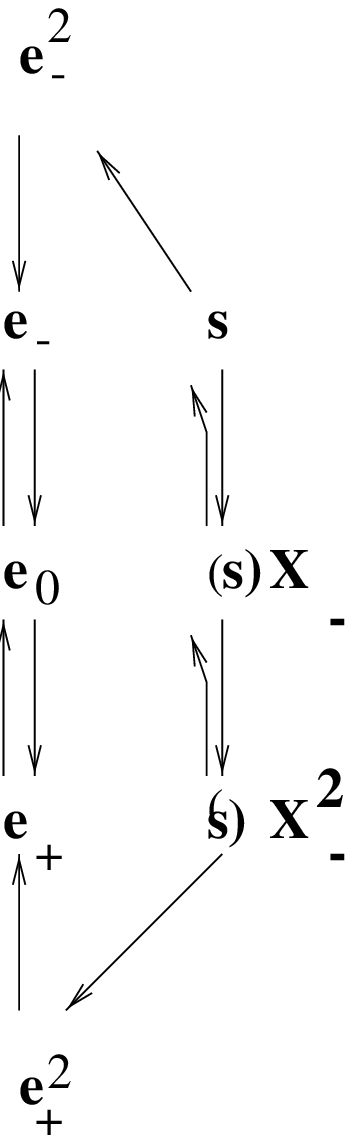} }
\bigskip

 Using the formulae for the right action of $U_q$ 
 we find that $(e_-^2) X_+  = (q^3 - q) e_- $. 
 Note that there is no $e_-^2e_+$ appearing in 
$ (e_-^2) X_+$. This means that $(e_-^2) X_+X_- = 0 $
 as indicated in Fig.3 by the absence of a upward arrow 
 emerging from the state at the second row. 
 To get the form of the element  $s$ of the $A_q$,  which 
 sits at the right end of the second row of Fig. 3
 we solve the equation $ ( s ) X_+ = e_-^2$. 
 This allows solutions 
\eqn\eqfors{ s = a e_- - e_-^3e_+^2 - q( 1 - q^2 )e_-^2e_+, }
 where $a$ is an arbitrary constant. Explicit expressions for the 
 other states on the right leg of Fig. 3 can be obtained  
 by acting with $X_+$ on the element $s$. 

The rep. $I_0^4$ is shown below. 
\bigskip
\fig\qfrsec{ 
{Fig.4 }  } 
{\epsfxsize1.0in\epsfbox{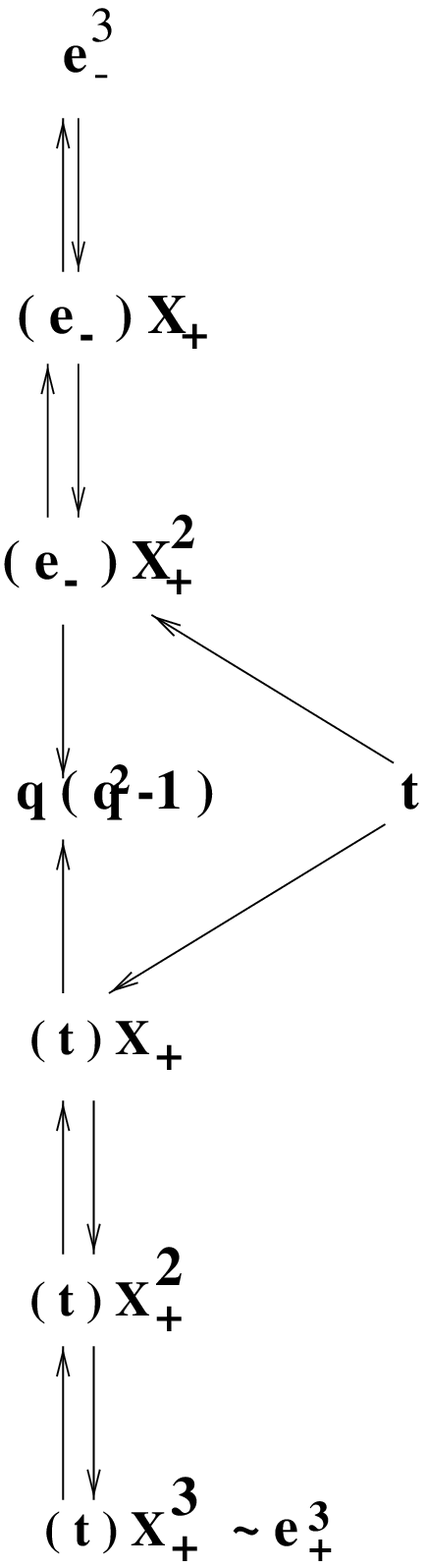} }
\bigskip

 Similarly we find that 
 $(e_3) X_+^2$ is proportional to the identity, 
 with no combination of $e_-e_+, e_-^2e_+^2, e_-^3e_+^3$.  
 And we can solve for the element $t$ up to an arbitrary
 constant $b$ by requiring that $ (t) X_-$  gives 
 $ (e_-^3) X_+^2$. This gives an expression
\eqn\tilds{ t  = b - qe_-e_+ + {1\over 2}e_-^2e_+^2 +
 { q(1+q^2)\over 2 }e_-^3e_+^3. }  
 
 The steps, described above for $q= e^{ i \pi \over 3}$
 and $q = e^{ i \pi \over 4 } $ make it clear how to obtain explicit 
 expressions form for polynomials which fill out the 
 appropriate set of I-reps in the case of general $M$.

\subsec{ Relation to fuzzy sphere } 

 The decomposition \mtwo\ contains as a sub-module 
 the direct sum of standard representations of $U_q$. 
 These are representations which have the same structure 
 as representations of ordinary $SU(2)$. For $U_q$ at roots of unity 
 there is a finite set of these, with spins 
 $ 2s \le M-2 $. This set of representations forms a closed fusion 
 ring and is used as a model for fusion rules 
 of WZW models \wzwqg. Integer spin representations
 within this range appear in the decomposition of the fuzzy sphere 
 \dec.  

 The representation $I_0^{p} $ contain as a sub-module 
 the representation $V_s$ with spin given by $2s = M-p$. 
 This means that included in  \mtwo\ is a direct sum 
 of the standard representations with integer spins ranging 
 up to $M-2$. This will allow us to exhibit some interesting
 properties of a map $ \rho : \Aq \rightarrow \Af ( { M\over 2 }  )  $ 
in the next
 section. 
 The same property holds for the case of odd $M$, i.e 
 we have a sub-module in \modtw\ which is the direct 
 sum of standard reps. with integer spins ranging up 
 to $( M-3) / 2 $. In this case we have a map $ \rho : 
 \Aq \rightarrow \Af ( { M-1 \over 2 }  )$.

\newsec{ Deformation Map,  deformed product and  deformed co-product } 

   The decomposition of $A_q$ contains as a submodule
   the $ { \cal H} = \oplus V_s $, where $V_s$ is the standard representation 
   of spin $s$. 
   While $\cH$ is a sub-module it is not a subalgebra 
   of $A_q$. It turns out that we can define a new product 
   on $\cH$, which we call $\mu_q^*:  \cH \times \cH \rightarrow \cH$, 
   and which is a natural additional product to consider on any 
   $U_q$ module algebra given the existence of twistings of the
   standard co-product of $U_q$. This new product will allow the
   sub-module to be, in addition, a sub-algebra and  will in fact be
   the same as the fuzzy sphere product. We will begin by elaborating
   the properties of the map $\rho : \Aq \rightarrow \Af $ 
  in connection with deformation maps, and
   then show the relation between the new product and twisted
   co-products. 

   We have a vector space isomorphism between 
   $ \cH $ and $\Af $. 
   Let us call this map $\rho : \Af \rightarrow \Aq $. 
   There is deforming map : 
\eqn\defuuq{ D : U \rightarrow U_q }
 Let the map $\lambda_f : A_f \otimes U \rightarrow A_f $ 
 denote the right action of the universal enveloping algebra of 
 $SU(2)$ on the fuzzy sphere. Let the map 
 $\lambda_f : A_q \otimes U_q \rightarrow A_q$ denote the right action
   of $U_q$ on the $q$-sphere. 
  The deforming map satisfies the property 
 \eqn\dfcond{ \rho \circ \lambda_f 
  =  \lambda_q \circ  ( \rho \otimes D ) }
 
 This is illustrated diagrammatically in the figure below. 

\bigskip 
\fig\qfrsec{ 
{Fig.5 }  } 
{\epsfxsize2.0in\epsfbox{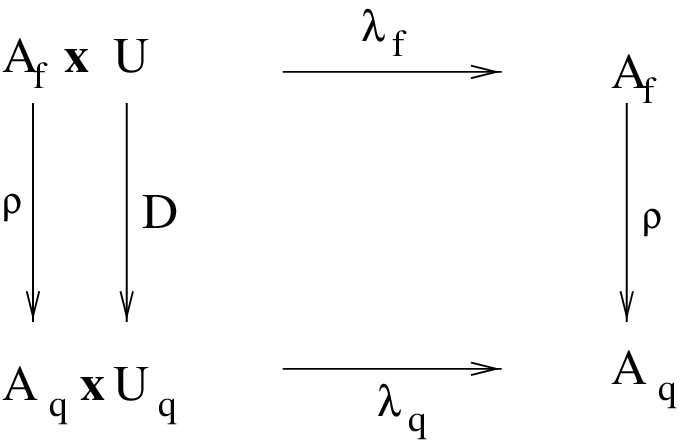} }
\bigskip
To be more explicit we can choose the isomorphism 
 $ \rho$ and the deformation map as follows. 
A highest weight of $U_q$ with $H$ eigenvalue $l$ is given by 
 $e_-^l$. A highest weight of $U$ with eigenvalue of $H$ equal to 
 $l$ is given by $S_-^l$. We define 
 $Y^{(q)}_{l,m} = \lambda_q ( e_-^l , X_+^{(l-m)} )$. 
 We also define $ Y^{(f)}_{l,m} = \lambda_f ( S_-^l , X_+^{(l-m)}) $ . 
 With these definitions $\rho$ takes a simple form 
 $\rho ( Y^{(f)} ) = Y^{(q)} $. And the deformation map 
 is : 
\eqn\deform{\eqalign{  
& K = q^{H} \cr 
& X_+^{(q)} = X_+ \cr 
& X_-^{(q)} = { ( l- H  ) ( l+ H + 1 ) \over  ( l- H  )_q ( l+ H + 1
 )_q } X_- \cr }} 
  
Here $l$ is understood to be expressed in terms
 of the generators of the algebra using 
 $l(l+1) = H^2 + {1 \over 2 } ( X_+X_- + X_-X_+ ) $. 
 If we use this formula 
in a space where $l$ takes the value   $M/2$, the denominator
 can vanish with the numerator finite. In the application of
 interest the eigenvalues of  $l$  extend
 from $0$ to $M/2 - 1 $ ( for $M$ even ) 
  and to $(M-1)/2 - 1 $ (for $M$ odd ). 

The content of \dfcond\ can now  be expressed 
 more simply 
 as 
\eqn\simpex{ \lambda_f ( Y^{(f)},  X ) = \lambda_{q} ( Y^{(q)}, D
  (  X ) ) } 
 The meaning of \dfcond\ and \simpex\ is that using  
 the action of $U_q$ on the  vector-subspace $\cH$ of $\Aq$
 we can reconstruct the action of the $U$ on $\Af$. So far 
 we have only discussed the module structure of $\cH$. We now turn to the
 product structure $\cH$. 

 We can define a new product on the $Y^{(q)} $ by first mapping 
 with $\rho$ and then multiplying. 
 The modified $q$-product is acted on by 
 the $q$-symmetry through the co-product : 
\eqn\copdmod{ 
 \Delta_q^* =  ( D \otimes D  ) \circ \Delta \circ D^{-1}
 }
 This may be seen as follows.
\eqn\showdelst{\eqalign{  \lambda_q ( Y^{(q)} * Y^{(q)} , X^{(q)} ) 
&= ( \lambda_f \otimes \lambda_f ) ( Y^{(q)} \otimes Y^{(q)}, \Delta
\circ D^{-1}
(X^{(q)} ) )   \cr 
& = ( \lambda_q \otimes \lambda_q ) ( Y^{(f)} \otimes Y^{(f)} , (D
\otimes D ) \circ \Delta \circ D ( X^{(q)} ) ) \cr }} 

This new co-product for the $q$-algebra 
 can in fact be written as a conjugation of the 
standard co-product by an element $F$ of $ U_q \otimes U_q$. 
 The existence of such an element for large $M$, follows 
 from work of Drinfeld \drinf.  
 The work of \cgz\ shows how to construct 
 it from Clebsch-Gordan coefficients of the $q$-symmetry 
 and the classical symmetry. The twist element is related to the 
 $R$ matrix but unlike the $R$ matrix it is not very explicitly 
 known \cc.  Drinfeld twists have recently appeared in discussions
 of brane world-volume non-commutativity recently \oeck\wat.
 Some of their abstract properties are discussed in generality 
 in \gz.

 The existence of this twist of the standard 
 co-product of $U_q$ allows the  definition of  a new product 
 on the sub-module $\cH $ of the $A_q$ module algebra. This new product 
 allows the sub-module to  be, in addition,  a sub-algebra.
 The action of $U_q$ on the sub-module with multiplication 
 on the sub-module defined by the star product recovers
 the fuzzy sphere and its $U(SU(2)$ symmetry. 
 This transformation was possible because of the fact observed 
 in section 4 that $\Aq$ contains a sub-module which transforms 
 under $U_q$ exactly the way $\Af$ trasnforms under the classical 
 symmetry. We expect that these observations 
 will allow an understanding of how to relate field theories
 naturally written in terms of $q$-sphere variables
to field theories written in terms of fuzzy sphere variables. 
 These transformations may have analogies to
 the Seiberg-Witten map \sw.

\newsec{ Summary and Outlook } 

 We have elaborated on connections between different kinds of
 non-commutative spheres. We gave a 
 strategy for recovering the 
 fuzzy 2-sphere from the $q$-sphere at roots of unity. 
 Our main result relates to  the module structure 
 of  $\Aq$ and its relation via deformation maps 
 to the module structure and the product structure 
 of the fuzzy sphere. Further work is needed to understand 
 the detailed implications for maps between field theories 
 defined on $q$-sphere and field theories defined 
 on fuzzy sphere. A preliminary remark is that the subset 
 of theories
 defined on $\Aq$ which  only use the sub-module  made of $V$ reps
 and use the deformation product, can be mapped to fuzzy sphere field
 theories. It remains to study in more detail issues of reality,  
 invariant traces and field theoretic Feynman rules 
  in the light of the deformation maps and Drinfeld
 twists. Some works that are likely to be useful in this direction 
 are \oeck. 

 One intriguing fact we have uncovered about the fuzzy sphere 
 in this investigation is that $\Af ( M/2  ) $ and $\Af ( M )$ ( for M
 even ) are both 
 module algebras which have the same finite dimensional symmetry 
 algebra ( the finite $U_q$ algebra for $q = e^{ i \pi \over M}$).  
  $ \Af (M) $ is a module algebra with the standard $q$-co-product. 
 $\Af (M/2)$ is a module algebra with a twisted $q$-co-product.

  We already discussed two physical motivations 
  for this work in the introduction.  
  Another direction where work along these lines 
     can be useful is towards the formulation of 
  a precise relation between the structure of the chiral ring 
  of theories dual to string theory on $ AdS \times S $
 and quantum group symmetries. In some sense the chiral ring 
 of the orbifold CFT dual to $ AdS_3 \times S^3$ would be analogous 
 to the $q$-sphere algebra. It would be a module algebra which admits 
 a left and right action of $U_q(SU(2))$.  We might also look for 
 action of $U_q(SU(1,1)) $ but even if such an action exists, 
 it would be simpler to first focus of $SU(1,1)$ highest weights and
 identify the $U_q(SU(2))$ action.   
  The existence of some 
 relations between the fuzzy sphere and the chiral ring
 \jmrii\ and the results of the current paper relating fuzzy sphere 
 to quantum groups give good reason to expect that a lot 
 of information about the chiral ring might be encoded in the 
 existence of a hidden quantum group symmetry.

 There are other connections between fuzzy spheres
 and quantum group symmetric spaces 
 explored in \stein. These $q$ symmetric spaces are  module 
 algebras ( with the standard $q$-co-product )  
 which involve the $V$-representations of \keller. 
 The finite dimensional quotient we studied has 
 $I$-reps which in turn contain the $V$-reps sub-module. 
 By considering  conjugations of the standard 
 co-product of $U_q$ by the deformation map, 
 we were lead to define a new product on the sub-module 
 which makes it a sub-algebra as well. 
 It will be interesting to see if there are examples 
 of finite dimensional $U_q$ module algebras
 where one has mixtures of $I$ and $V$ reps ( 
 when we use the standard $q$-coproduct )  and which can nevertheless
 be related to fuzzy spheres after an appropriate modification of the 
 product based on deformation maps.

 It will be interesting to look for generalization 
 of the connections between q-2sphere and fuzzy 2-sphere 
 to the case of 4-spheres. $q$-4-spheres can be constructed 
 along the lines of \sugi. At roots of unity we expect finite
 dimensional truncations to exist. 
  Relations between the q-sphere and its differential 
 calculi with  the fuzzy 4-sphere \csw\ would be a good test 
 of usefulness of the non-commutative spheres in providing  models of 
 non-commutative space-time where  similarities in the  
 finite $N$ physics of 
 different backgrounds  entering the
 ADS/CFT correspondence can be made manifest. 
 The need for considering 
 differential calculi would appear necessary  from the observation of 
 \pmstu\ that the deformed algebras relevant in the fuzzy  4-sphere 
 case mix momenta and coordinates. 
 Similar questions can be asked about the relation between classical 
 and quantum spaces for the $AdS$ part of space-time. 
 Some relevant works on non-compact quantm groups are 
 \steinck\suz\roch\tesh\liho\dob.

  There have been other  recent appearances of fuzzy spheres in the
  literature \recfuz. 
  It will be very interesting to explore whether  a 
  picture of non-commutativity of space-time in string theory,
  e.g along the lines of \yoney\jeyo,
  can coherently account for these diverse
 appearances of fuzzy structures.

\bigskip
 
 \noindent{\bf Acknowledgements:}
 We are especially grateful to Phillippe Roche 
 for several illuminating discussions near the 
 beginning of this project. We have also benefitted from 
 useful communications with P.M. Ho, A. Matusis, J. Pawelczyk, 
  H. Steinacker, L. Susskind. This research was supported by DOE grant  
 DE-FG02/19ER40688-(Task A).

\listrefs

\end